\documentclass{emulateapj}

\shorttitle{Star--forming galaxies}
\shortauthors{de Mello et al.}

\begin{document}


\title{Star--forming galaxies at intermediate redshifts: morphology, ages and sizes\\
    }


\author{D. F. de Mello\altaffilmark{1,2,3}, Y. Wadadekar\altaffilmark{4}, T. Dahlen\altaffilmark{5}, S.
Casertano\altaffilmark{4}, J. P. Gardner\altaffilmark{1}}

\altaffiltext{1}{Observational Cosmology Laboratory, Code 665, Goddard Space Flight Center, Greenbelt, MD
20771}
\altaffiltext{2}{Catholic University of America Washington, DC 20064}

\altaffiltext{3}{Johns Hopkins University, Baltimore, MD 21218}

\altaffiltext{4}{Space Telescope Science Institute, Baltimore, MD 21218}

\altaffiltext{5}{Department of Physics, Stockholm University, SE-106 91 Stockholm, Sweden}



\begin{abstract}

We present the analysis of the deepest near-UV image obtained with 
Hubble Space Telescope
using the WFPC2(F300W) as part of the parallel observations of the Hubble Ultra
Deep Field campaign. The U-band 10$\sigma$ limiting magnitude measured over 0.2 arcsec$^{2}$ is 
m$_{\rm AB}$=27.5 which is 0.5 magnitudes deeper 
than that in the Hubble Deep Field North. 
We matched the U-band catalog with those in the ACS
images (B, V, i, z) taken during the Great Observatories Origins Deep Survey observations of the 
Chandra Deep Field South and obtained photometric redshifts for 306 matched objects. 
We find that the UV-selected galaxies span all the major morphological types at 0.2 $<$$z_{\rm phot}$$<$ 1.2. 
However, disks are more common at lower redshifts, 0.2 $<$$z_{\rm phot}$$<$ 0.8. 
Higher redshift objects (0.7 $<$$z_{\rm phot}$$<$ 1.2) are on average bluer than lower$-z$ and 
have spectral type typical of starbursts. Their morphologies are compact, peculiar or 
low surface brightness galaxies. Despite the UV-selection, 13 objects have spectral types of early-type galaxies; two of
them are spheroids with blue cores. The evolutionary synthesis code, Starburst99,
was used to age-date the UV-selected galaxies which were found to have rest-frame colors 
typical of stellar populations with intermediate ages $>$ 100 Myr. The average half-light radius (rest-frame  1200--1800 \AA)
of the UV-selected galaxies at 0.66$<$$z_{\rm phot}$$<$1.5 is $0.26 \pm 0.01$ arcsec ($2.07  \pm 0.08$ kpc). 
The UV-selected galaxies are on average fainter (M$_{\rm B}$=--18.43$\pm$0.13) than 
Lyman Break Galaxies (M$_{\rm B}$=--23$\pm$1). Our sample includes early-type galaxies that are
presumably massive and forming stars only in their cores, as well as
starburst-type systems that are more similar to the LBGs, although
much less luminous. This implies that even the starbursts in our
sample are either much less massive than LBGs or are forming stars at
a much lower rate or both. The low surface brightness galaxies have no
overlap with the LBGs and form an interesting new class of their own.

\end{abstract}





\section{Introduction}

Establishing how galaxies formed and evolved to become today's
galaxies remains one of the fundamental goals of theorists and
observers. The fact that we see a snapshot of the universe as if it
was frozen in time, prevents us from directly following the process of
galaxy assembly, growth, ageing, and morphological metamorphosis with
time. The alternative commonly pursued is to look for evolutionary
signatures in surveys of large areas of the sky .  Recently, Heavens
et al. (2004) analyzed the `fossil record' of the current stellar
populations of $\sim$100,000 galaxies ($0.005<z<0.34$) from the Sloan
Digital Sky Survey (SDSS) and noted a mass dependence on the peak
redshift of star--formation. They claim that galaxies with masses
comparable to a present-day L* galaxy appears to have experienced a
peak in activity at $z\sim0.8$. Objects of lower (present-day stellar)
masses ($< 3
\times 10^{11}$M$_{\odot}$) peaked at $z\le$0.5. Bell et al. (2004)
using the COMBO-17 survey (Classifying Objects by Medium-Band
Observations in 17 filters) found an increase in stellar mass of the
red galaxies (i.e. early--types) by a factor of two since $z\sim
1$. Papovich et al. (2005) using the HDF-N/NICMOS data suggest an
increase in the diversification of stellar populations by $z\sim$1
which implies that merger--induced starbursts occur less frequently
than at higher redshifts, and more quiescent modes of star-formation
become the dominant mechanism. Simultaneously, around $z\sim$1.4, the
emergence of the Hubble--sequence galaxies seems to occur.

Connecting the star formation in the distant universe ($ z > 2$) to
that estimated from lower redshift surveys, however, is still a
challenge in modern astronomy. Using the Lyman break technique
(e.g. Steidel et al. 1995), large samples of star--forming galaxies at
$2<z<4.5$ have been identified and studied. Finding unobscured star forming
galaxies in the intermediate redshift range ($0.5<z<1.5$) is more difficult since the
UV light ($\lambda$ $\sim$  1000--2000 \AA) that comes from young and massive OB stars
is redshifted into the near-UV. The near-UV detectors are less
sensitive than optical ones which makes UV imaging expensive in telescope time. 
For instance, $\sim$30\% of
HST time in the Hubble Deep Field campaign was dedicated to the U-band
(F300W - $\lambda_{\rm max}$ = 2920 \AA), whereas the other 70\% was shared between B, V, and
I-bands. Inspite of this, the limiting depth reached in the U band is
about a magnitude shallower than in the other bands. 

Recently, Heckman et al. (2005) attempted to identify and study the
local equivalents of Lyman break galaxies using images from the
UV-satellite GALEX and spectroscopy from the SDSS. 
Amongst the UV
luminous population, they found two kinds of objects: 1) massive
galaxies that have been forming stars over a Hubble time which
typically show morphologies of late-type spirals; 2) compact galaxies
with identical properties to the Lyman break galaxy population at
$z\sim 3$. These latter are genuine starburst systems that have formed
the bulk of their stars within the last 1--2 Gyr.

Establishing the population of objects that contributes to the
rise in the SFR with lookback time has strong implications to theories
of galaxy evolution and can only be confirmed by a proper census of
the galaxy population at the intermediate$-z$ epoch ($0.4<z<1.5$).  In
the present paper we identify a sample of intermediate redshift UV
luminous galaxies and seek to understand their role in galaxy
evolution.  We have used data from the Great Observatories Origins
Deep Survey (GOODS) in combination with an ultra deep UV image taken
with HST/WFPC2 (F300W) to search for star-forming galaxies.  The
space-UV is the ideal wavelength to detect unobscured star-forming
galaxies whereas the multiwavelength ACS images (B, V, i, z) are ideal
for morphological analysis of the star-forming objects.

This paper is organized as follows: \S 2 describes the data processing, \S 3 presents the sample, 
\S 4 discusses redshifts, \S 5 presents various issues 
concerning their colors and age, \S 6 describes the morphological
classification, \S 7 discusses the sizes while and presents
comparison with Lyman Break Galaxies. Finally, \S 8 summarizes the main 
conclusions. Throughout this paper, we use a cosmology
with $\Omega_{\rm M}=0.3$, $\Omega_{\Lambda}=0.7$~and $h=0.7$.
Magnitudes are given in the AB-system.

\section{The Data}

\begin{figure*}
\epsscale{.9}
\plotone{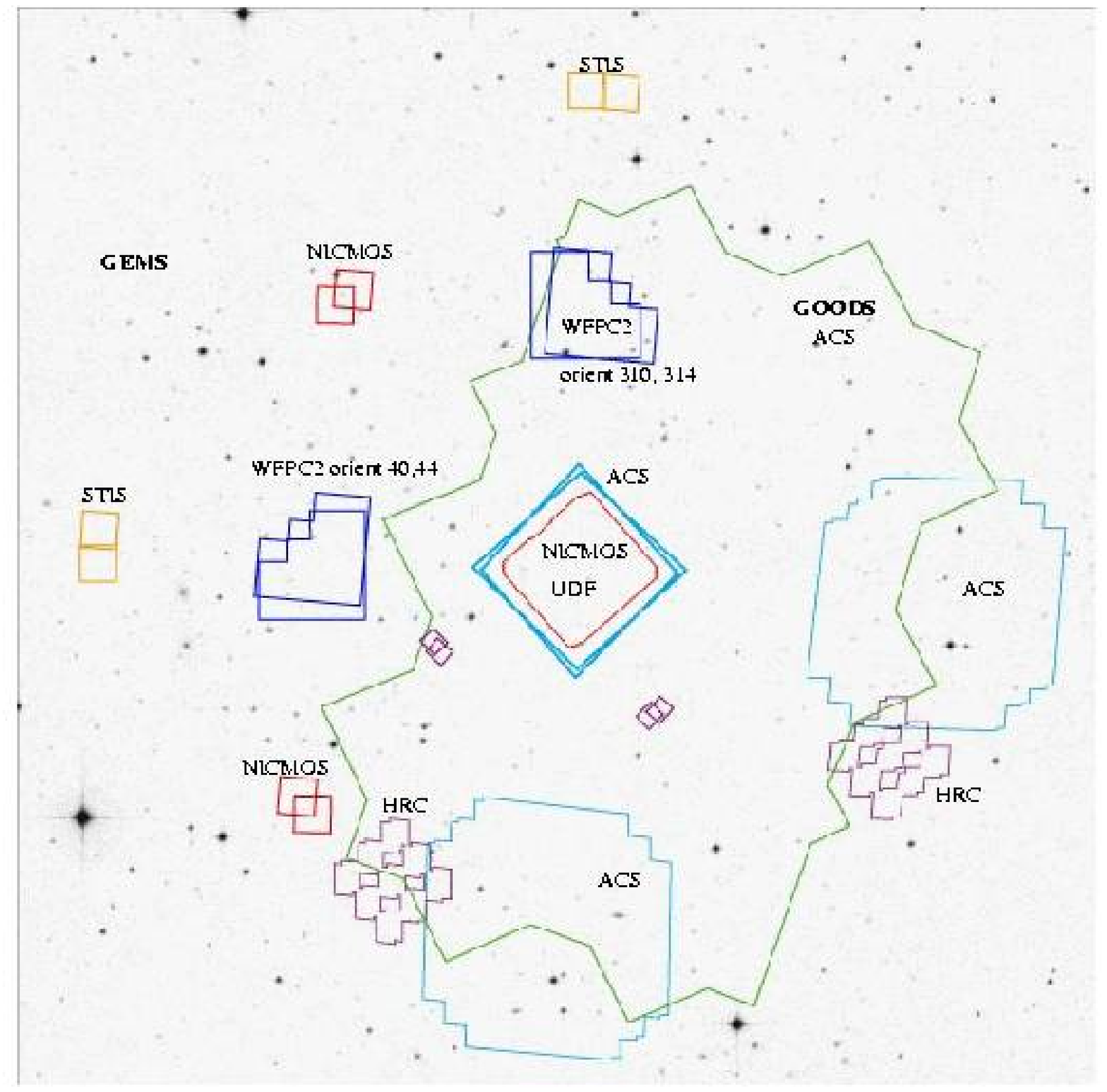}
\epsscale{1}
\caption{The Chandra Deep Field South region observed by the Great Observatories Origins Deep Survey 
(GOODS - in green)  and by the Ultra Deep Field (UDF) campaigns with the Advanced Camera for Surveys
(ACS - in blue). During the ACS and the Near Infrared Camera and Multi-Object Spectrometer (NICMOS - in red) 
observations of the UDF, all other HST instruments obtained data in parallel configuration.
In this article, we are analyzing the WFPC2 orient 310 and 314 data obtained with the F300W filter.   
\label{hudfpar2w}}
\end{figure*}

The Ultra Deep Field (UDF) provided the deepest look at the universe
with HST taking advantage of the large improvement in sensitivity in
the red filters that ACS provides. In parallel to the ACS UDF other
instruments aboard HST also obtained deep images
(Fig. \ref{hudfpar2w}). In this paper we analyze the portion of the
data taken with the WFPC2 (F300W) which falls within the GOODS-S area
(Orient 310/314); another WFPC2 image overlaps with the Galaxy
Evolution From Morphology and SEDs (GEMS) survey area.  Each field
includes several hundred exposures with a total exposure time of 323.1
ks and 278.9 ks respectively. The 10$\sigma$ limiting magnitude
measured over 0.2 arcsec$^{2}$ is 27.5 magnitudes over most of the
field, which is about 0.5 magnitudes deeper than the F300W image in
the HDF-N and 0.7 magnitudes deeper than that in the HDF-S.

\subsection{Data Processing}

A total of 409 WFPC2/F300W parallel images, with exposure times
ranging from 700 seconds to 900 seconds overlap partially with the GOODS-S
survey area. Each of the datasets was obtained at one of two
orientations of the telescope: (i) 304 images were obtained at Orient
314 and (ii) 105 images were obtained at Orient 310.

We downloaded all 409 datasets from the MAST data archive along with
the corresponding data quality files and flat fields. By adapting the
drizzle based techniques developed for data processing by the WFPC2
Archival Parallels Project (Wadadekar et al. 2005), we constructed a
cosmic ray rejected, drizzled image with a pixel scale of 0.06
arcsec/pixel. Small errors in the nominal WCS of each individual image
in the drizzle stack were corrected for by matching up to 4 star
positions in that image with respect to a reference image.

Our drizzled image was then accurately registered with respect to the
GOODS images by matching sources in our image with the corresponding
sources in the GOODS data, which were binned from their original scale
of 0.03 arcsec/pixel to 0.06 arcsec/pixel. Once the offsets between
the WFPC2 image and the GOODS image had been measured, all 409 images
were drizzled through again taking the offsets into account, so that
the final image was accurately aligned with the GOODS images.

The WFPC2 CCDs have a small but significant charge transfer efficiency
problem (CTE) which causes some signal to be lost when charge is
transferred down the chip during readout. The extent of the CTE
problem is a function of target counts, background light and
epoch. Low background images (such as those in the F300W filter) at
recent epochs are more severely affected. Not only sources, but also
cosmic rays leave a significant CTE trail. We attempted to flag the
CTE trails left by cosmic rays in the following manner: if a pixel was
flagged as a cosmic ray, adjacent pixels in the direction of readout
(along the Y-axis of the chip) were also flagged as cosmic-ray
affected. The number of pixels flagged depended on the position of the
cosmic ray on the CCD (higher row numbers had more pixels
flagged). With this approach, we were able to eliminate most of the 
artifacts caused by cosmic-rays in the final drizzled image.

\section{Catalogs}

We detected sources on the U-band image using SExtractor (SE) version
2.3.2 (Bertin \& Arnouts 1996).  Our detection criterion was that a
source must exceed a $1.5\sigma$ sky threshold in 12 contiguous pixels. We
provided the weight image (which is an inverse variance map) output by
the final drizzle process as a {\it MAP$_{-}$WEIGHT} image to
Sextractor with {\it WEIGHT$_{-}$TYPE} set to {\it
MAP$_{-}$WEIGHT}. This computation of the weight was made according to
the prescription of Casertano et al. (2000). It takes into account
contributions to the noise from the sky background, dark current, read
noise and the flatfield and thus correctly accounts for the varying
S/N over the image, due to different number of overlapping datasets at
each position. During source detection, the sky background was
computed locally. A total of 415 objects were identified by SE.

\begin{figure}
\plotone{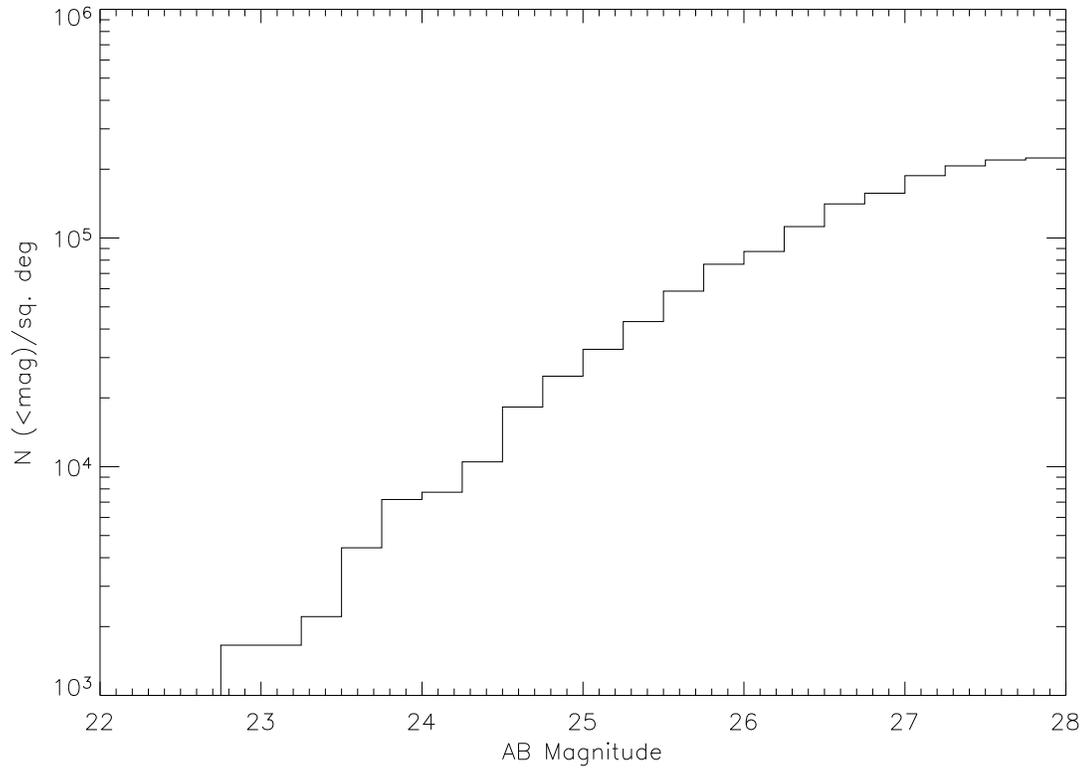}
\caption{
Cumulative galaxy counts from the HUDF F300W parallel observations of the
GOODS CDFS field. {\it MAG$_{-}$AUTO} magnitudes (F300W) from SExtractor were used in
obtaining the number counts.\label{numcounts}}
\end{figure}

Fig.~\ref{numcounts} shows the cumulative galaxy counts using {\it
MAG$_{-}$AUTO} magnitudes (F300W) from SE. Only sources within the
region of the image where we have full depth data were included in
this computation.

\section{Redshifts}

Spectroscopic redshifts are available for 12 of the objects in the
F300W catalog (taken from the ESO/GOODS-CDFS spectroscopy master
catalog\footnote{
http://www.eso.org/science/goods/spectroscopy/CDFS$_{-}$Mastercat/}).
For the remaining objects, we calculate photometric redshifts using a
version of the template fitting method described in detail in Dahlen
et al. (2005).  The template SEDs used cover spectral types E, Sbc,
Scd and Im (Coleman et al.  1980, with extension into UV and NIR-bands
by Bolzonella et al. 2000), and two starburst templates (Kinney et
al. 1996).

In addition to data from the F300W band, we use multi-band photometry
for the GOODS-S field, from $U$~to $K_s$~bands, obtained with both
$HST$~and ground-based facilities (Giavalisco et al. 2004). As our
primary photometric catalog, we use an ESO/VLT ISAAC $K_s$-selected
catalog including $HST$~WFPC2 $F300W$~and ACS $BViz$~data, combined
with ISAAC $JHK_s$ data. We choose this combination as our primary
catalog due to the depth of the data and the importance to cover both
optical and NIR-bands when calculating photometric redshifts. This
catalog provides redshifts for 72 of the objects detected in the
$F300W$~band. The two main reasons for this relatively low number is that
part of the WFPC2 {\it $F300W$}~image lies outside the area covered by
ACS+ISAAC, and that
UV selected objects are typically blue and may therefore be too faint to be
included in a NIR selected catalog. For these objects, we use a ground-based photometric
catalog selected in the $R$-band which includes ESO (2.2m WFI,
VLT-FORS1, NTT-SOFI) and CTIO (4m telescope) observations covering
$UBVRIJHK_s$. This adds 146 photometric redshifts. Finally, to derive
photometric redshifts for objects that are too faint for inclusion in
either of the two catalogs described above, we use ACS $BViz$ and
WFPC2 $F300W$ photometry to obtain photometric redshifts. This adds 76
photometric redshifts to our catalog. In summary, we have
spectroscopic redshifts for 12 objects and photometric redshifts for
294. Subsequent analysis
in this paper, only includes the 306 sources with photometric or
spectroscopic redshifts.

\begin{figure}
\plotone{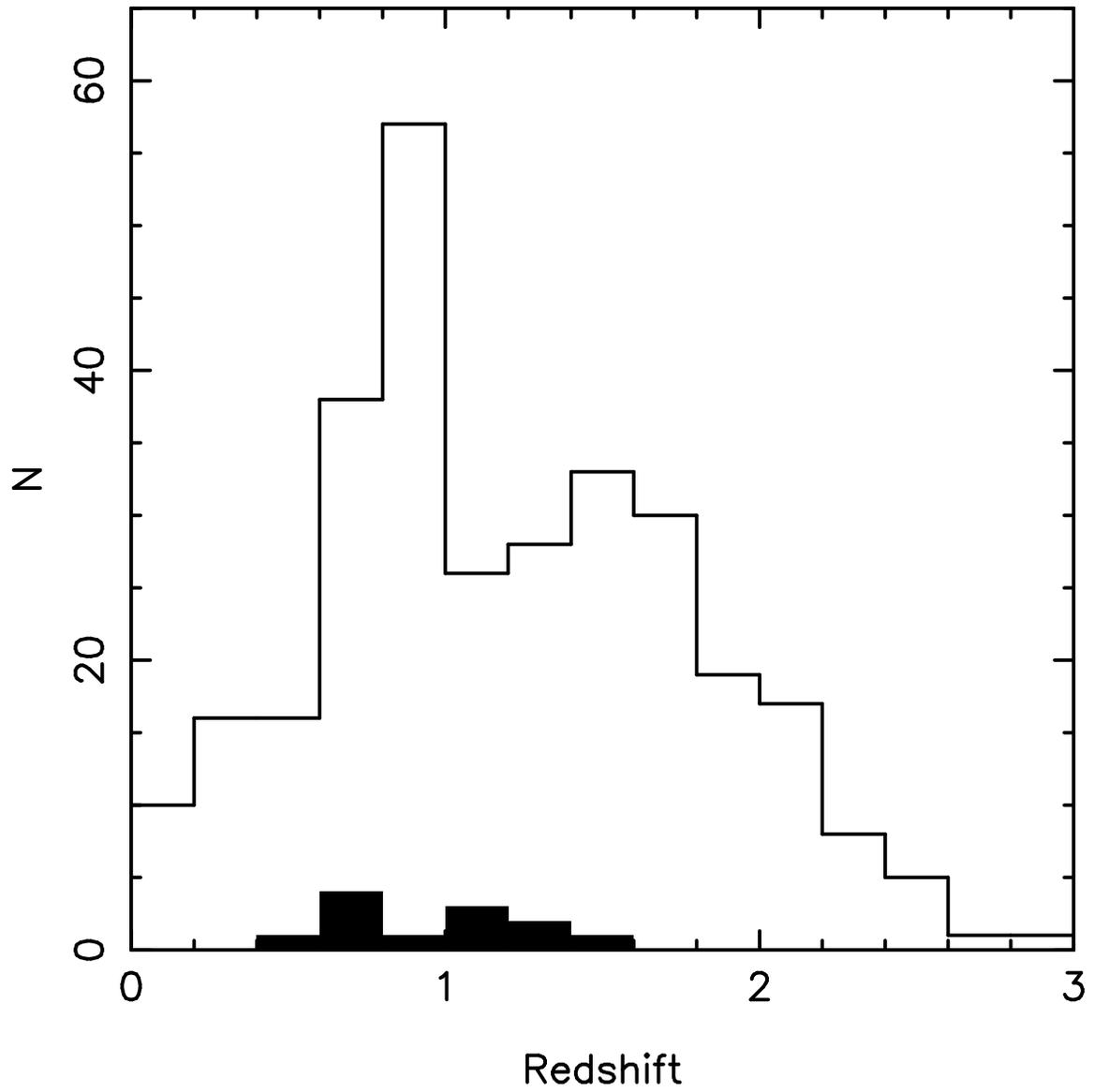}
\caption{Redshift distribution for 306 of the total 415 objects detected in F300W.
Black histogram shows spectroscopic redshifts. The remaining are 
photometric redshifts.
\label{histphtzall}}
\end{figure}

The remaining 109 objects in the $F300W$~catalog belong to one or more
of the following four categories (i) outside the GOODS coverage area
(ii) too faint for photometric redshifts to be determined, (iii)
identified as stars (iv) are `single' objects in the optical (and/or
NIR) bands but are fragmented into multiple detections in the
$F300W$-band. In such cases, photometric redshifts are only calculated
for the `main' object.
The redshift distribution of our sample is shown in Figure \ref{histphtzall}. 

To investigate the redshift accuracy of the GOODS method, we compare the
photometric redshifts with a sample of 510 spectroscopic redshifts taken
from the ESO/GOODS-CDFS spectroscopy master catalog. We find an overall
accuracy $\Delta_z\equiv\langle|z_{\rm phot}-z_{\rm spec}|/(1+z_{\rm spec})\rangle\sim 0.08$ 
after removing a small fraction ($\sim$3\%) of outliers with $\Delta_z>0.3$.
Since starburst galaxies, which constitute a large fraction of our sample,
have more featureless spectra compared to earlier type galaxies with a
pronounced 4000\AA-break, we expect the photometric redshift accuracy to
depend on galaxy type. Dividing our sample into starburst and non-starburst 
populations, we find $\Delta_z\sim$0.11 and $\Delta_z\sim$0.07, respectively.
This shows that the photometric redshifts for starburst have a higher scatter,
the increase is, however, not dramatic. Also, the distribution of the residuals
(spectroscopic redshift -- photometric redshift), has mean value that is
close to zero for both, the starburst and the total population. Therefore,
derived properties such as mean absolute magnitudes and mean rest-frame colors,
should not be biased due to the photometric redshift uncertainty.

\section{Colors}

\begin{figure}
\plotone{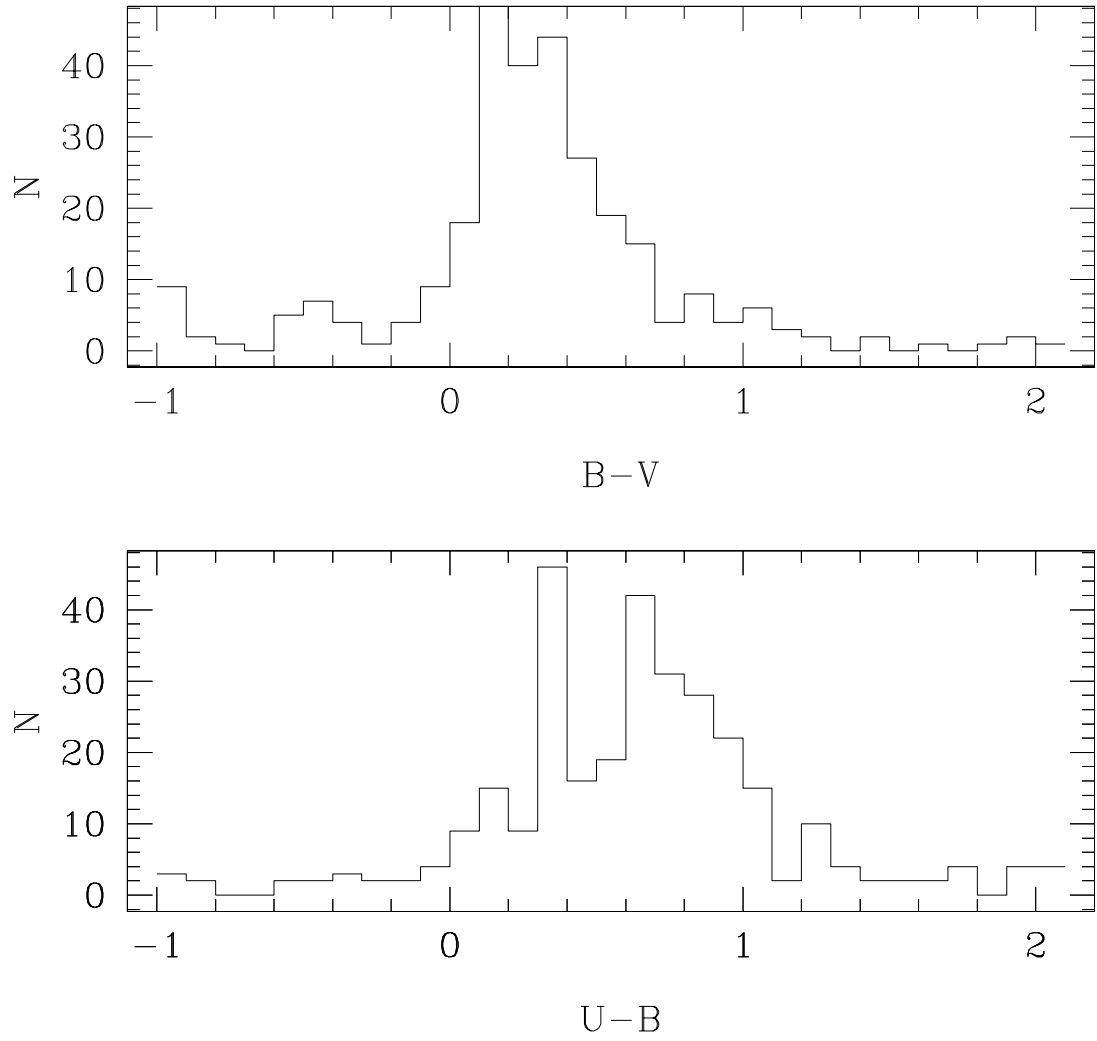}
\caption{Rest-frame U--B and B--V color distribution for all UV-selected galaxies. \label{histub}}
\end{figure}

\begin{figure}
\plotone{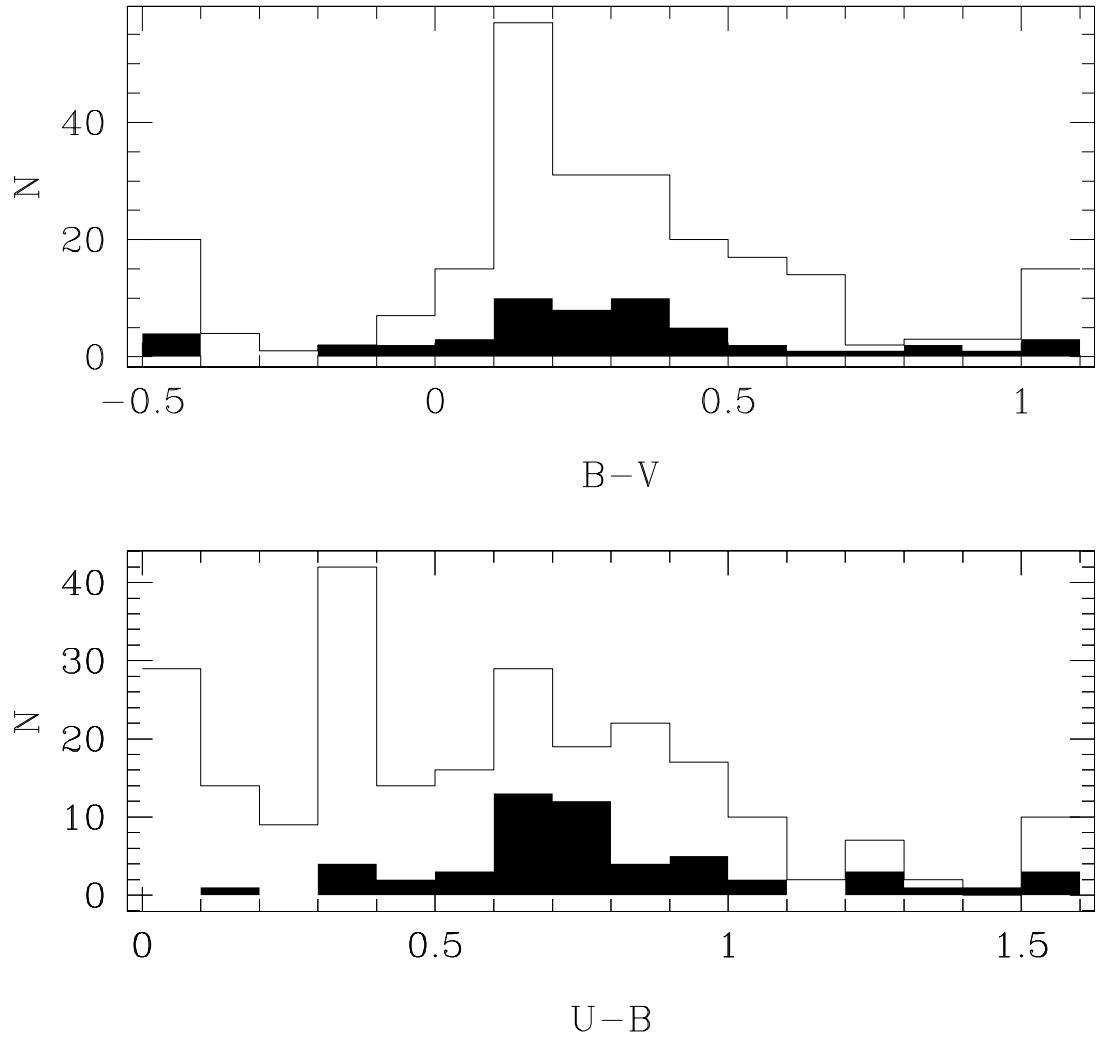}
\caption{U--B and B--V color distribututions. Objects with $z_{\rm phot}<0.7$ are in shaded histograms and
objects with $z_{\rm phot}>0.7$ are in the unshaded histograms. \label{histubz}}
\end{figure}

Using information from the photometric redshifts, rest-frame absolute 
magnitudes and colors are calculated using the recipe in Dahlen et al. (2005).
 The rest-frame U--B and B--V color distributions (Fig.~\ref{histub}) show a peak in the blue side of
the distribution (U--B$\sim$0.4 and B--V$\sim$0.1). The majority of
the objects that have these colors are actually in the high redshift
bin and have $z_{\rm phot} > 0.7$ as shown in Fig.~\ref{histubz}. The bimodality in colors seen in the HDF-S
(Wiegert, de Mello \& Horellou 2004) is not seen in this sample which
is UV-selected and deficient in red objects.

\begin{figure}
\plotone{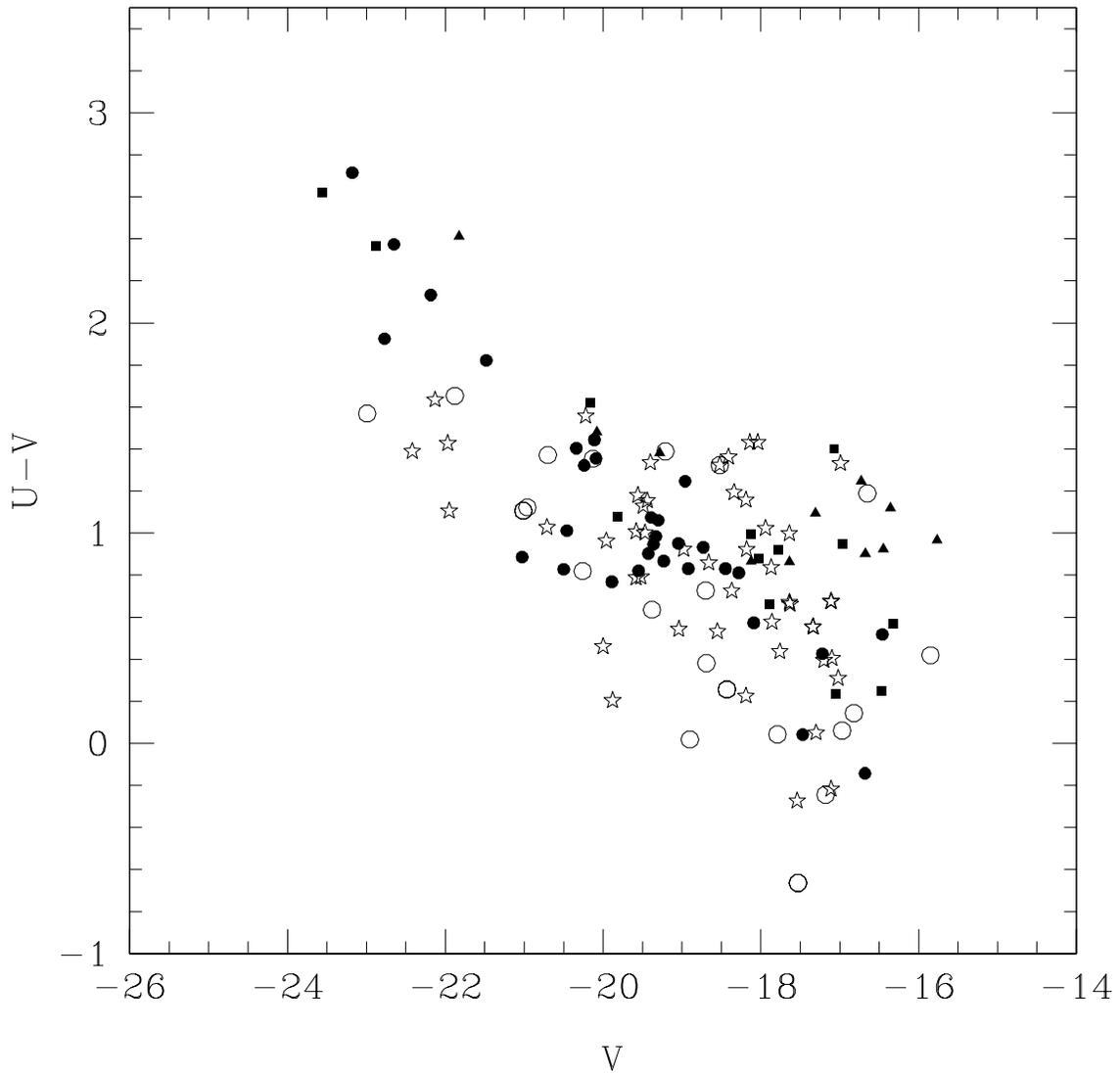}
\caption{Rest-frame U-V color against absolute magnitude V. Symbols are: 
filled triangles=$0.2<z<0.4$, filled squares=$0.4<z<0.6$, filled circles=$0.6<z<0.8$, stars=$0.8<z<1.0$, empty circles=$1.0<z<1.2$.
 \label{plotuvvphotz1p2}}
\end{figure}

In Fig.~\ref{plotuvvphotz1p2}, we show the rest-frame U--V color and V-band absolute
magnitude of all galaxies with $0.2<z_{\rm phot}<1.2$. The trend is similar to
the one found recently by Bell et al. (2005) for $\sim$1,500 optically-selected  $0.65 \le z_{\rm phot}<0.75$ 
galaxies using the 24 $\mu$m data from the Spitzer Space Telescope in combination with COMBO-17, GEMS
and GOODS. 
However, the 25 galaxies in our UV-selected sample, which are in the same redshift range, are on 
average redder (U--V=0.79 $\pm$ 0.13 
(median=0.83)) and fainter (M$_{\rm V}$=--19.1 $\pm$ 0.32 (median=--19.3))
than the average values in Bell et al. of all visually-classified types. This is due to
the low depth of the GEMS survey coverage (one HST orbit per ACS pointing) which was used to 
provide the rest-frame V-band data of their sample. 
The UV-selected galaxies we are analyzing have deeper GOODS multiwavelength data (3, 2.5, 2.5 and 5 orbits per 
pointing in B, V, i, z, respectively) which
GEMS lacks whenever outside the GOODS field.

\begin{figure}
\plotone{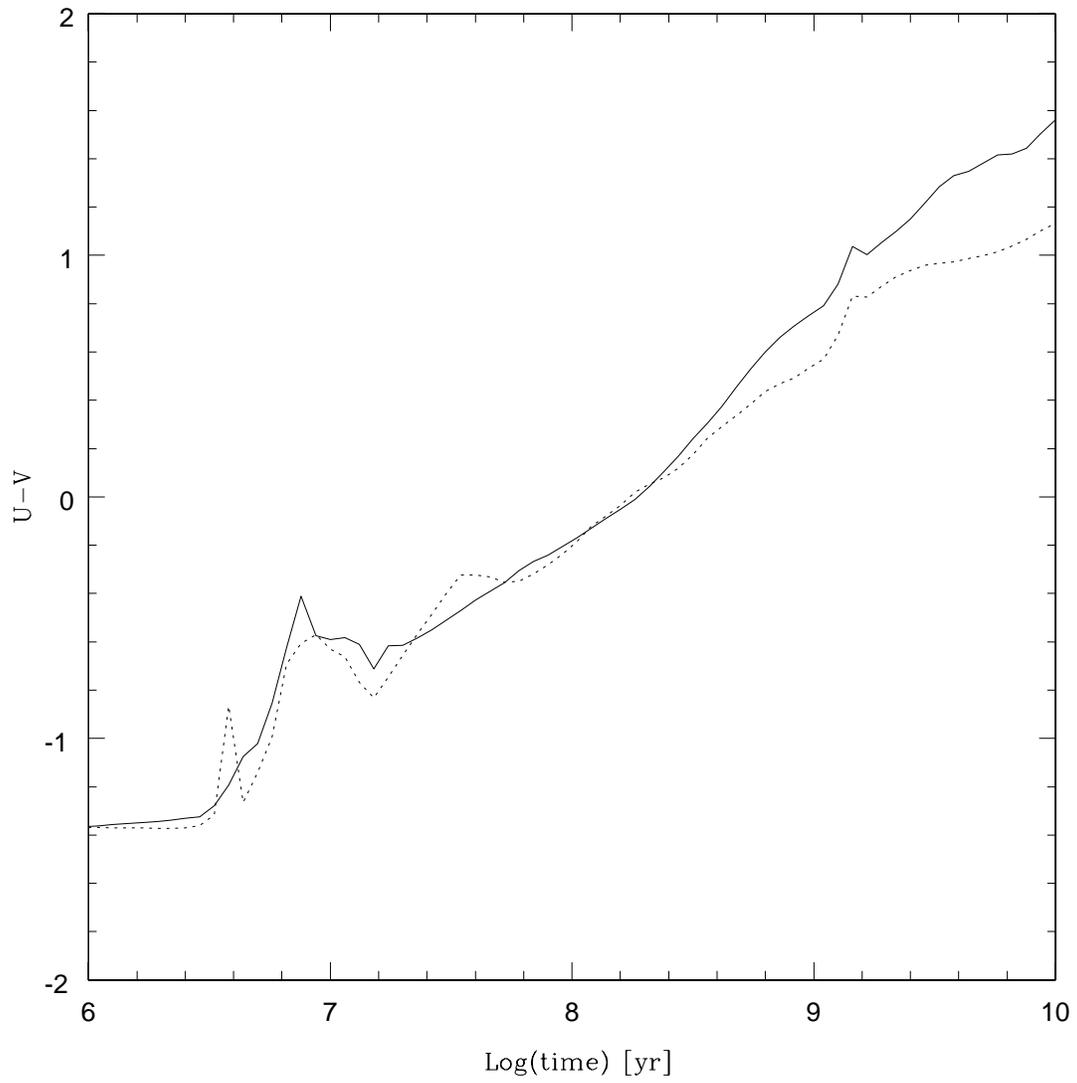}
\caption{U-V color evolution from Starburst99 (Vazquez \& Leitherer 2005) - solid line: solar
metallicity, dotted line: sub-solar metallicity (20\% solar).
\label{plotuvagesb99}}
\end{figure}

Fig.~\ref{plotuvagesb99} shows the U--V color evolution produced using the
new version of the evolutionary synthesis code, Starburst99 (Vazquez \& Leitherer 2005)
with no extinction correction. The new code (version 5.0) is optimized to reproduce all stellar phases 
that contribute to the integrated light of a stellar population from young to old ages. 
As seen from Fig.~\ref{plotuvagesb99}, the UV-selected sample has U--V colors
typical of ages $>$100 Myr (U--V $>$ 0.3; average U--V=0.79$\pm$0.06). 
The 25 objects with $0.65 \le z_{\rm phot}<0.75$, for example, 
have U--V typical of ages 10$^{8.4}$ to 10$^{10}$ yr. Although we cannot
rule out that these object might have had a different star formation history, 
and not necessarily produced stars continuously as adopted in the model shown, they do not have 
the U--V colors of young instantaneous bursts (10$^{6}$ yr) which have typically U--V $<$ --1.0 (Leitherer et al. 1999).

Vazquez \& Leitherer (2005) have tested the predicted colors by comparing the models to sets of observational
data. In Fig.~\ref{plotvibvdatasb99} we reproduce their Fig.~19, a color-color plot of the super star clusters and 
globular clusters of NGC 4038/39 (The Antennae) by Whitmore et al. (1999) together with model predictions and our data 
of UV-selected galaxies. No reddening correction was applied to the clusters which can
be as high as E(B-V)=0.3 due to significant internal reddening in NGC 4038/49. The clusters are divided into
three distinct age groups (i) young, (ii) intermediate ages (0.25 -- 1 Gyr) and (iii) old (10 Gyr). 
Vazquez \& Leitherer analyzed the effects of age and metallicity in the color predictions and
concluded that age-metallicity degeneracy in the intermediate-age range ($\sim$ 200 Myr) is not a
strong effect. This is the age when the first Asymptotic Giant Branch (AGB) stars influence the colors in their models. 
The vertical  loop at (B--V)$\sim$ 0.0-0.3 is stronger at solar metallicity and is caused by Red Super Giants which are much less
important at lower abundances. We interpret the large spread in the color--color plot of our sample 
as a combination of age, metallicity and extinction correction. The latter can bring some of the outliers closer 
to the model predictions, e.g. an E(B-V)=0.12 running parallel to the direction of metallicity and age evolution 
would bring more objects closer to the younger clusters with ages $<$ 0.25 Gyr.

\section{Morphology}

Classifying the morphology of faint galaxies has proved to be a very
difficult task (e.g. Abraham et al. 1996; van den Bergh et al. 1996;
Corbin et al. 2001; Menanteau et al. 2001) and automatized methods are
still being tested (e.g. Conselice 2003, Lotz et al. 2005). In such a
situation, spectral types which are obtained from the template fitting
in the photometric redshift technique are a good morphology indicator
(e.g. Wiegert, de Mello \& Horellou 2004) and in combination with
other indicators help constrain galaxies properties. In
Fig.~\ref{histst} we show the distribution of the spectral types of our
sample. As expected in a UV-selected sample, the majority of the
objects have SEDs typical of late-type and starburst galaxies. This
trend does not uniformly hold if we separate the sample in redshift
bins (Fig.~\ref{histstz}). The lower redshift bin ($z_{\rm phot}<0.7$)
has a mix of all types whereas the higher redshift bin has mostly
($\sim$60\%) starbursts.

The average absolute magnitudes for the different spectral types in
the UV-selected sample are M$_{\rm B}$= --20.59 $\pm$ 0.24 (E/Sa),
M$_{\rm B}$= --18.61 $\pm$ 0.17 (Sb-Sd-Im), M$_{\rm B}$= --17.80 $\pm$
0.16 (Starbursts). The median absolute magnitudes for these types of
galaxies are M$_{\rm B}$= --20.52 (E/Sa), --18.71 (Sb-Sd-Im) and
--17.62 (Starbursts) which are, except for the early-types, fainter
than the GOODS-S sample M$_{\rm B}$ = --20.6 (E/Sa), --19.9 (Sb-Sd),
and --19.6 (starburst) (Mobasher et al. 2004). This difference is due
to the magnitude limit (R$_{\rm AB}$ $<$ 24) imposed in that sample
selection, which was not used in our UV-selected sample; i.e. our
UV-selected sample is probing fainter objects at the same redshift
range ($0.2 < z_{\rm phot} < 1.3$). Despite the fact that our sample
is UV-selected, there are 13 objects with SEDs typical of early-type
galaxies (E/Sa) at this redshift range. Two of them are clearly
spheroids with blue cores ($z_{\rm phot}$$\sim$0.6--0.7,
B--V$\sim$0.7--0.8 and B$\sim$--22) and are similar to the objects
analyzed recently in Menanteau et al. (2005). These objects are
particularly important since they can harbor a possible connection
between AGN and star-formation.

Studies of the HDF-N has shown how difficult it is, to interpret galaxy
morphology at optical wavelengths, when they are sampling the rest
frame UV for objects at high redshifts. In the rest-frame near-UV
galaxies show fragmented morphology, i.e. the star-formation that
dominates the near-UV flux is not constant over the galaxy, but occurs
in clumps and patchy regions (Teplitz et al. 2005). Therefore,
rest-frame optical wavelengths give a better picture of the structure
and morphology of the galaxies. We used the ACS (BVi) images to
visually classify our sample and adopted the following
classification: (1) elliptical/spheroid, (2) disk, (3) peculiar, (4) compact,
(5) low surface brightness, (6) no ACS counterpart. Objects classified
as compact have a clear nuclear region with many showing a tadpole 
morphology; objects classified as peculiar are either interacting systems or
have irregular morphologies; objects classified as
low-surface-brightness (lsb) do not show any bright nuclear region,
and objects classified as (6) are outside the GOODS/ACS image. The
distribution of types as a function of redshift is shown in
Fig.~\ref{histmorph} and reveals two interesting trends: (i) the
decrease in the number of disks at $z>0.8$ and (ii) the increase in
the number of compact and lsb galaxies at $z_{\rm phot}>0.8$. 
Moreover, as seen in Fig.~\ref{histmorphst}, there is a clear difference
in the morphology of starbursts (dashed line in the figure) and non-starbursts. 
Starbursts tend to be compact, peculiar or lsb while the non-starbursts have all
morphologies. 
Since our sample is UV-selected, star-forming disks are either less common
at higher$-z$ or there is a selection effect which is responsible for
the trend. For instance, we could have missed faint disks which hosts
nuclear starbursts and classified the object as compact. Deeper
optical images are needed in order to test this possibility. 

In Fig.~\ref{plotbbvall} we compare our sample properties of colors and
luminosity with typical objects from Bershady et al. (2000) which
includes typical Hubble types, dwarf ellipticals and luminous blue
compact galaxies at intermediate redshifts. Clearly, the UV-selected
sample has examples of all types of galaxies. However, a populated
region of the color-luminosity diagram with M$_{\rm B}$ $>$ --18 and
B--V$<$ 0.5 does not have counterparts either among the local Hubble
types or among luminous blue compact galaxies. The average morphology
of those objects is $4.21 \pm 0.58$ (type 4 is compact and type 5 is
lsb), 38\% are compact and 45\% are lsb, the remaining 17\% are either
spheroids or disks. 87\% of them have spectral types $>$ 4.33
(spectral types 4 and 5 are typical of Im and starbursts).

\section{Sizes}

We have used the half-light radii and the Petrosian radii to estimate the sizes of the
galaxies following the steps described in Ferguson et al. (2004). Half-light radius was measured
with SExtractor and the Petrosian radius was measured following the prescription adopted
by the Sloan Digital  Sky Survey (Stoughton et al. 2002). In order to 
estimate the overall size of galaxies, and not only the size of the star-forming region, 
we measured sizes as close to the rest-frame B band as possible, i.e. objects with 0.2$<$$z_{\rm phot}$$<$0.6 had
theirs sizes measured in the F606W image, objects with 0.6$<$$z_{\rm phot}$$<$0.8 in the F775W image,
and objects with 0.8$<$$z_{\rm phot}$$<$1.2 in the F850LP image. 
The correspondence between the two size measures was verified 
except for a few outliers: (i) three
objects with r$_{\rm h}$ $>$20 pixel (1 pixel = 0.06 arcsec/pixel) and
Petrosian radius $>$50 pixel which are large spirals, and (ii) an 
object with r$_{\rm h}$ $\sim$21 pixel and Petrosian radius $\sim$44
pixel which is a compact blue object very close to a low surface
brightness object. The half-light radius of the latter object is
over-estimated due to the proximity of the low surface brightness
object.

In Fig.~\ref{histlightarcsec} we show the observed half-light radii
(arcsec) distribution per redshift interval. The increase of small
objects at 0.8$<$$z_{\rm phot}$$<$1.0 is related to what is seen in
Fig.\ref{histmorph} where the number of compact galaxies peaks at the
same redshift interval, i.e. compacts have smaller sizes. The majority
of the objects at 0.8$<$$z_{\rm phot}$$<$1.2 have r$_{\rm h}$ $<$ 0.5
arcsec in the rest-frame B band. For comparison with high-$z$ samples 
which measure the sizes of galaxies at 1500 \AA, we measured the half-light radius in
the F300W images of all galaxies with 0.66$<$$z_{\rm phot}$$<$1.5,
corresponding to rest frame wavelength in the range 1200--1800
\AA. The average r$_{\rm h}$ is $0.26 \pm 0.01$ arcsec ($2.07  \pm 0.08$ kpc).

Fig.~\ref{plotblightkpchdf} shows the distribution of the derived
half-light radii (kpc) as a function of the rest-frame B
magnitudes. Five objects have r$_{\rm h}$ $>$ 10 kpc and are not
included in the figure. The broad range in size from relatively
compact systems with radii of 1.5--2 kpc to very larger galaxies with
radii of over 10 kpc agrees with the range in sizes of the luminous
UV-galaxies at the present epoch (Heckman et al. 2005). We included in
Fig.~\ref{plotblightkpchdf} the low--$z$ sample (0.7$<z<1.4$) from
Papovich et al. (2005) which is selected from a near-infrared,
flux-limited catalog of NICMOS data of the HDF-N. We have compared
r$_{\rm h}$ and M$_{\rm B}$ for the two samples, ours and Papovich et
al. (2005), using Kolmogorov-Smirnov (KS) statistics and found that
the UV-selected and the NIR-selected samples are not drawn from the
same distribution at the 98\% confidence level (D=0.24 and D=0.26 for r$_{\rm h}$ and M$_{\rm B}$,
respectively - D is the KS maximum vertical deviation between the two
samples).  The median values of the UV-selected objects is   r$_{\rm
h}$=3.02 $\pm$ 0.11 kpc and M$_{\rm B}$=--18.6 $\pm$ 0.1 which are
larger and fainter than the NIR-selected sample values of r$_{\rm h}$=
2.38 $\pm$ 0.06 kpc M$_{\rm B}$=--19.11 $\pm$ 0.07. This is due to a
number of low surface brightness objects (36\% or 16 out of 44) that
are found in our sample which are faint (M$_{\rm B}$ $> -20$) and
large (r$_{\rm h}$ $\ge$ 3 kpc).  These objects are not easily
detected in NIR but are common in our UV-selected sample due to the
depth of the U-band image which can pick up star-forming LSBs.

It is interesting to see how the properties of galaxies in our sample
compare with Lyman Break Galaxies at $2<z<4.5$. Despite the fact that
they are both UV-selected, LBGs belong to a class of more luminous
objects. Typical M$_{\rm B}$ of LBGs at $z\sim$3 are --23.0$\pm$1
(Pettini et al. 2001) whereas our sample has average M$_{\rm
B}$=--18.43$\pm$0.13. Three color composite images of the most
luminous objects in our sample (M$_{\rm B}$ $<$ --20.5) as shown in
Fig.~\ref{luminous}. There is clearly a wide diversity in morphology of
these objects. Four of them are clearly early-type galaxies, three are
disks showing either strong star formation or strong interaction, and
two of them are what we called low surface brightness and
compact. LBGs show a
wide variety in morphology ranging from relatively regular objects to
highly fragmented, diffuse and irregular ones. However, even the most
regular LBGs show no evidence of lying on the Hubble sequence. LBGs
are all relatively unobscured, vigourously star-forming galaxies that
have formed the bulk of their stars in the last 1-2 Gyr. Our sample is
clearly more varied: it includes early-type galaxies that are
presumably massive and forming stars only in their cores, as well as
starburst-type systems that are more similar to the LBGs, although
much less luminous. This implies that even the starbursts in our
sample are either much less massive than LBGs or are forming stars at
a much lower rate or both. The low surface brightness galaxies have no
overlap with the LBGs and form an interesting new class of their own.

\section{Summary}

We have identified 415 objects in the deepest near-UV image ever taken
with HST reaching magnitudes as faint as m$_{\rm AB}$=27.5 in the
F300W filter with WFPC2. We have used the GOODS multiwavelength images
(B, V, i, z) to analyze the properties of 306 objects for which we
have photometric redshifts (12 have spectroscopic redshifts). The main
results of our analysis are as follows:

\begin{enumerate}

\item UV-selected galaxies span all the major morphological types at 0.2 $<$$z_{\rm phot}$$<$ 1.2. 
However, disks are more common at lower redshifts, 0.2 $<$$z_{\rm phot}$$<$ 0.8.

\item Higher redshift objects (0.7 $<$$z_{\rm phot}$$<$ 1.2) are on average bluer than lower$-z$ and 
have spectral type typical of starbursts. Their morphologies are compact, peculiar or 
low surface brightness galaxies.

\item Despite of the UV-selection, 13 objects have spectral types of early-type galaxies; two of
them are spheroids with blue cores.

\item The majority of the UV-selected objects have rest-frame colors typical of stellar populations with 
intermediate ages $>$ 100 Myr.

\item UV-selected galaxies are on average larger and fainter than NIR-selected galaxies at 
0.7 $<$$z_{\rm phot}$$<$ 1.4; the majority of the objects are low-surface-brightness.

\item The UV-selected galaxies are on average fainter than Lyman Break Galaxies. 
The ten most luminous ones span all morphologies from early-types to low surface brightness. 

\end{enumerate}

\acknowledgments

\begin{figure}
\plotone{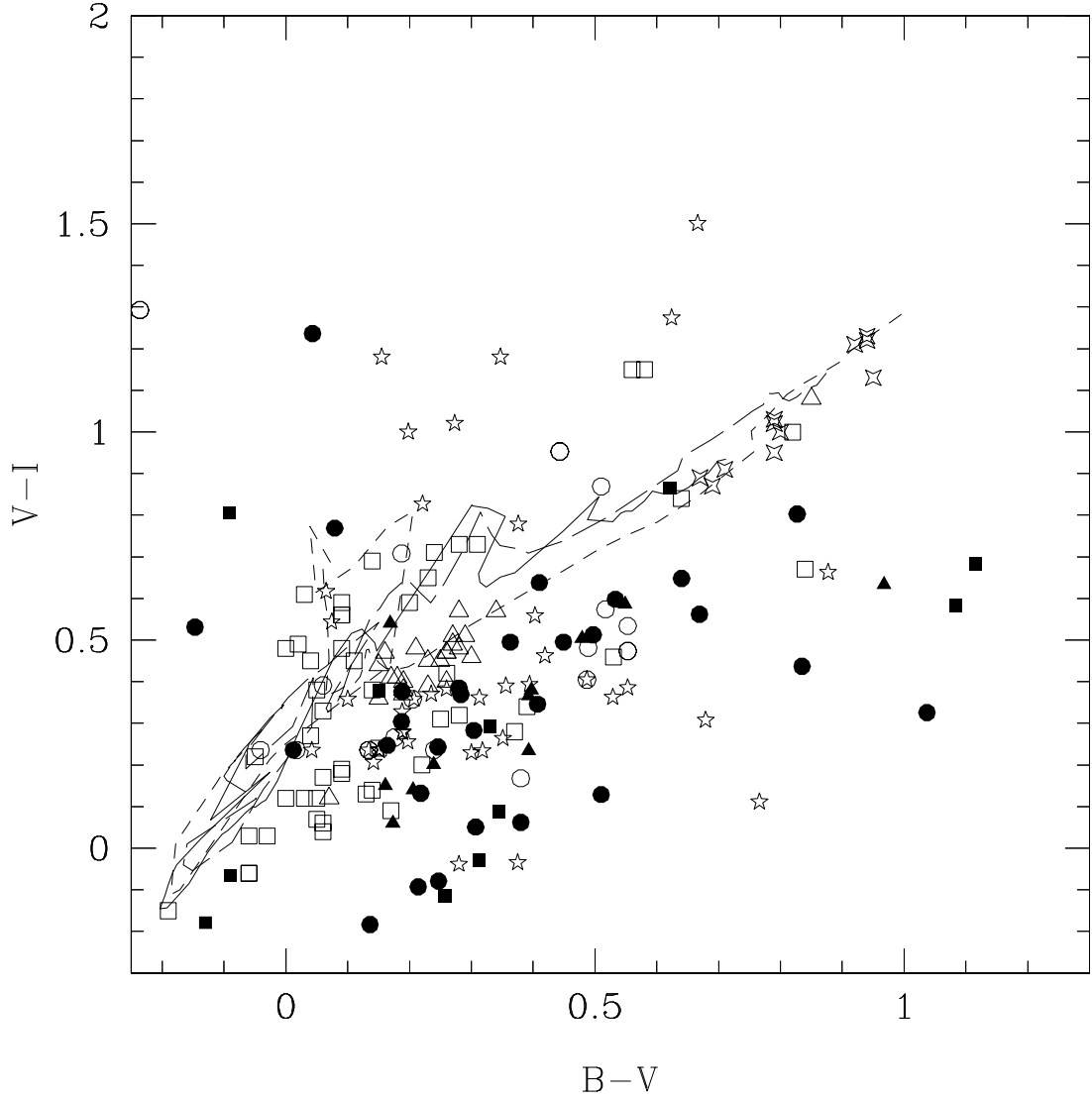}
\caption{Rest-frame  B--V against V--I. Symbols are: filled triangles=$0.2<z<0.4$, filled squares=$0.4<z<0.6$, 
filled circles=$0.6<z<0.8$, five pointed stars=$0.8<z<1.0$, empty circles=$1.0<z<1.2$, 
empty squares=young clusters
of NGC 4038/39 of Whitmore et al. (1999), empty triangles=intermediate ages, four pointed stars=old globular clusters.
Models are the same as in Vazquez \& Leitherer 2005 (Fig.19) for continuous star-formation, Salpeter IMF, 
up to 10 Gyr and for three metallicities: Z=0.0004 (solid line), Z=0.004 (long-dashed line) and
Z=0.02 (short-dashed line - solar). 
\label{plotvibvdatasb99}}
\end{figure}

\begin{figure}
\plotone{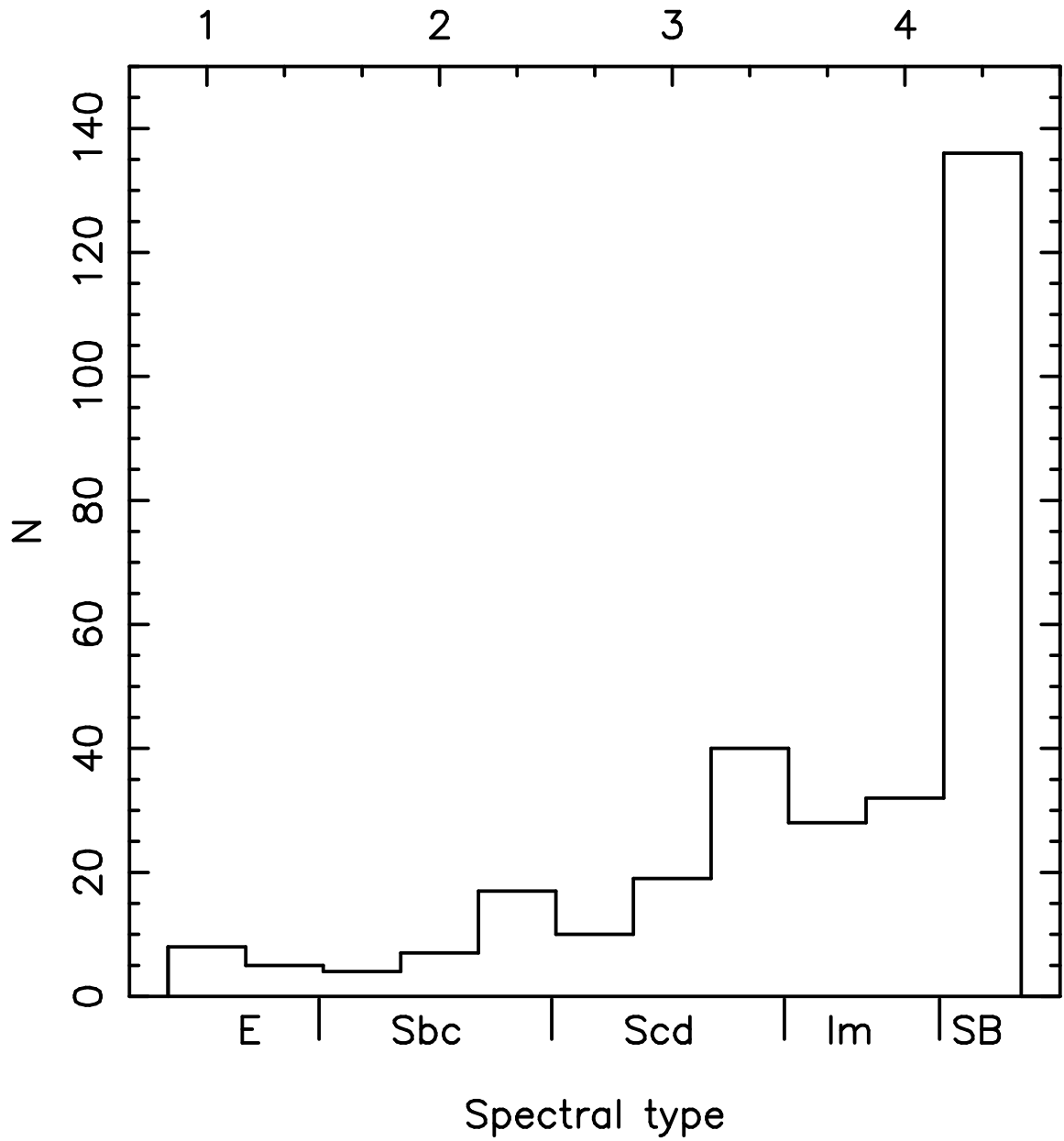}
\caption{Spectral type distribution. SED templates
E (1), Sbc (2), Scd (3) and Im (4) are from Coleman et al. (1980). Spectral type SB (5) 
corresponds
to starburst galaxies, i.e., either of the two starburst templates from 
Kinney et al. (1996).
 \label{histst}}
\end{figure}

\begin{figure}
\plotone{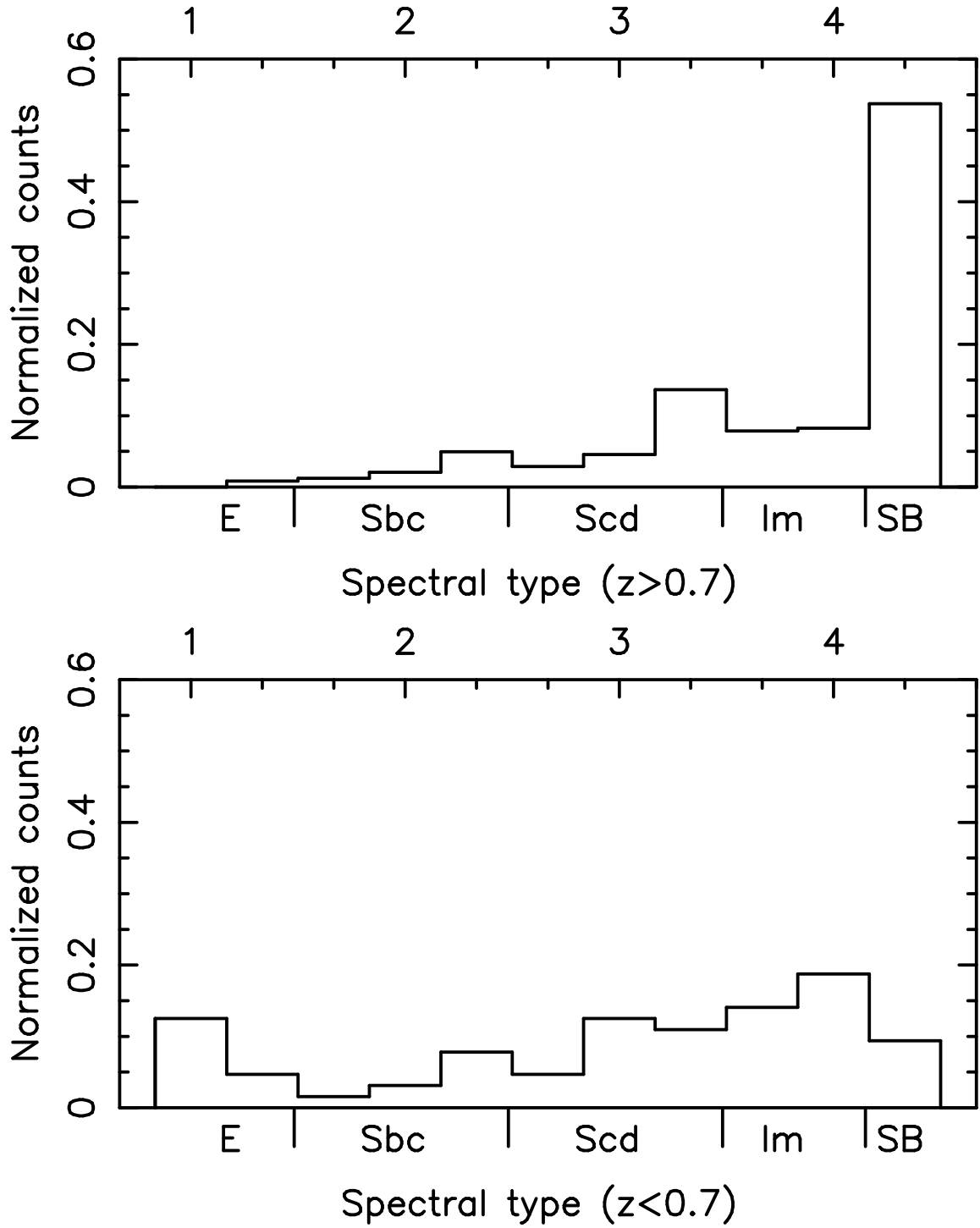}
\caption{Spectral type distribution for galaxies with redshift $z>0.7$ (top panel)
and $z<0.7$ (bottom panel). The numbers 1-4 correspond to the SED templates
E, Sbc, Scd and Im from Coleman et al. (1980). Spectral type $>4$ 
corresponds
to starburst galaxies, i.e., either of the two starburst templates from 
Kinney et al. (1996).
\label{histstz}}
\end{figure}

\begin{figure}
\plotone{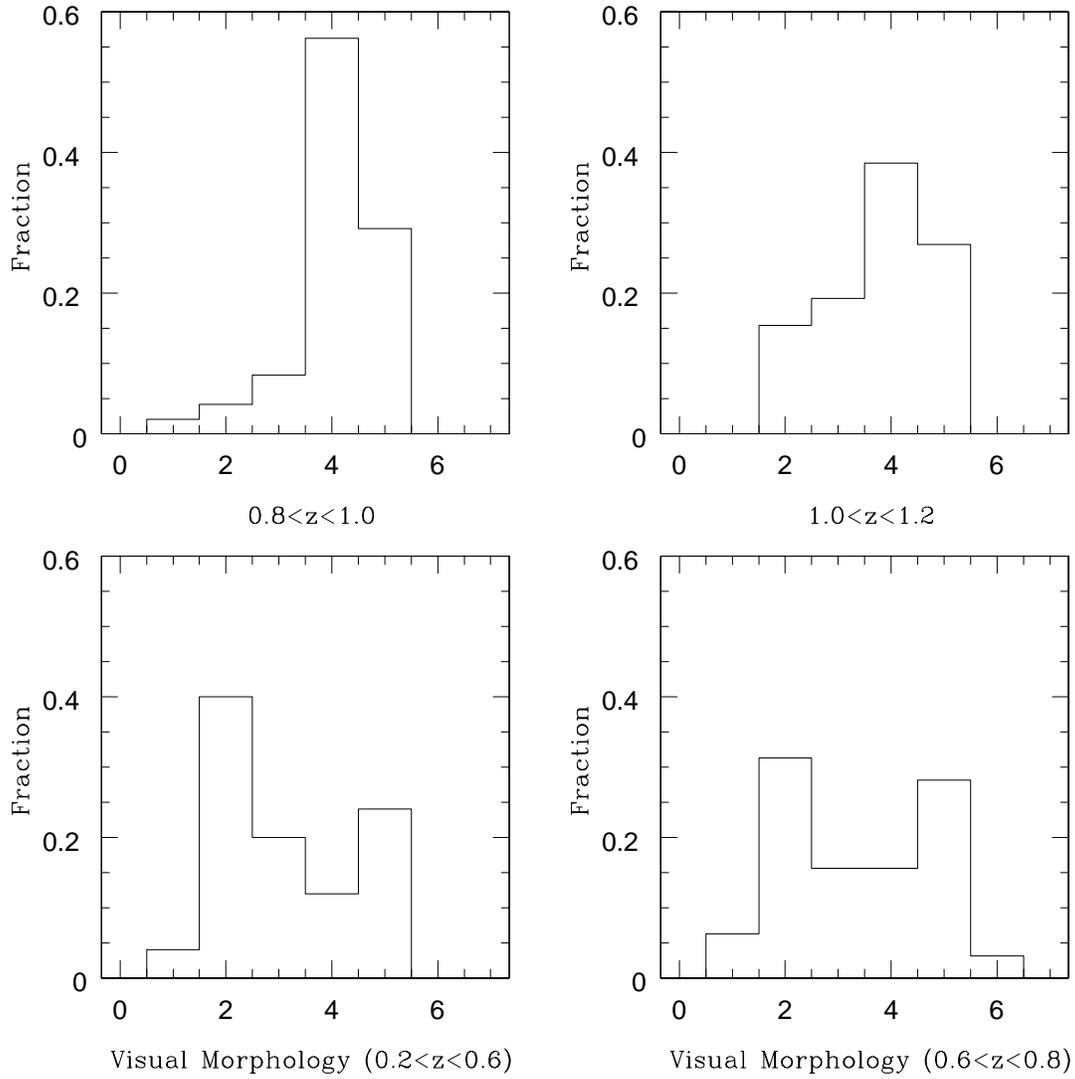}
\caption{Visual morphology distribution per redshift bin. Type 1=elliptical/spheroid, 2=disk, 3=peculiar, 4=compact, 5=low
surface brightness, 6=no ACS image.
\label{histmorph}}
\end{figure}

\begin{figure}
\plotone{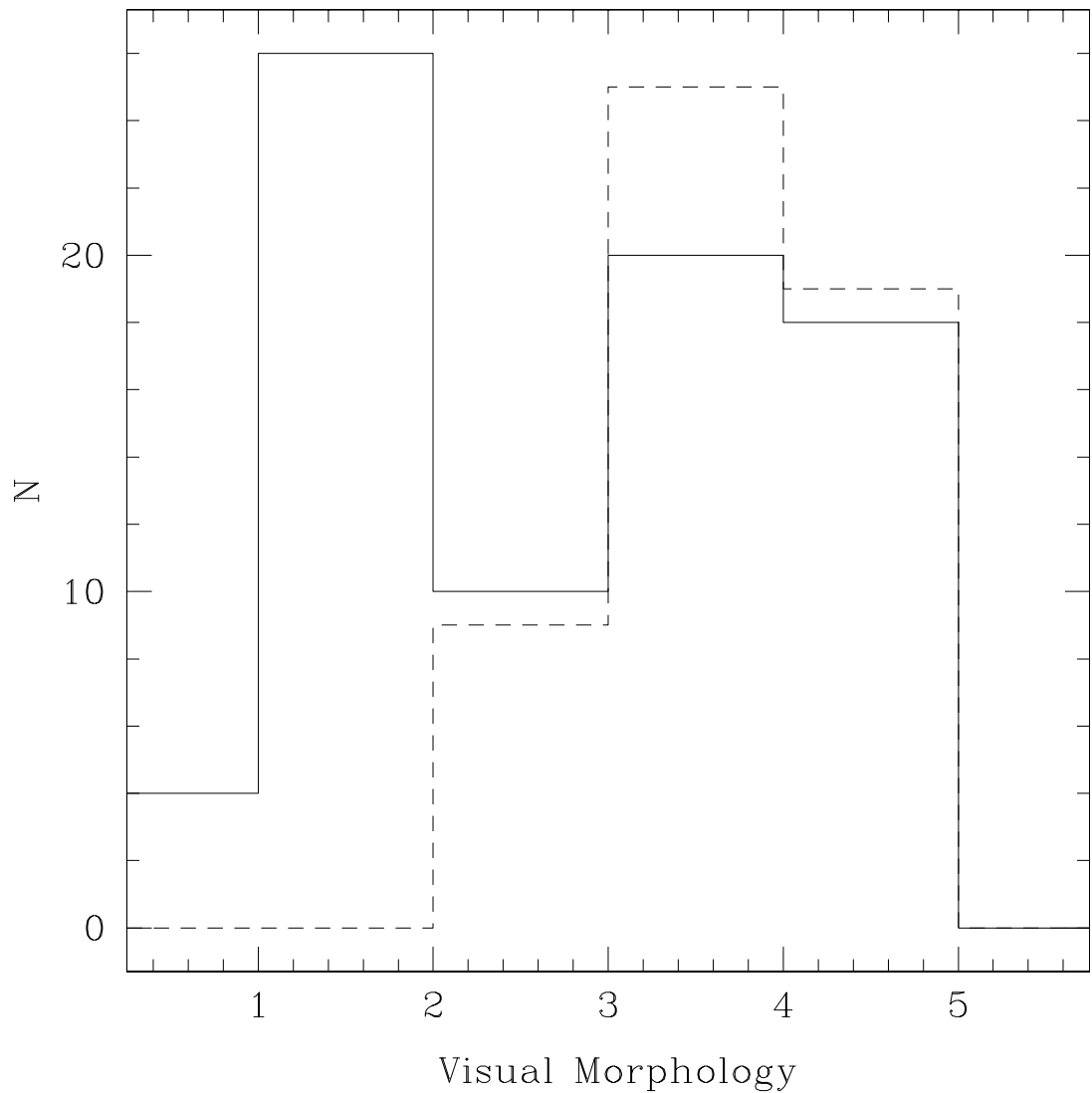}
\caption{Distribution of morphologies for starbursts (dashed) and non-starbursts (solid) with 
$z_{\rm phot}>0.7$. Type 1=elliptical/spheroid, 2=disk, 3=peculiar, 4=compact, 5=low surface brightness.
\label{histmorphst}}
\end{figure}

\begin{figure}
\plotone{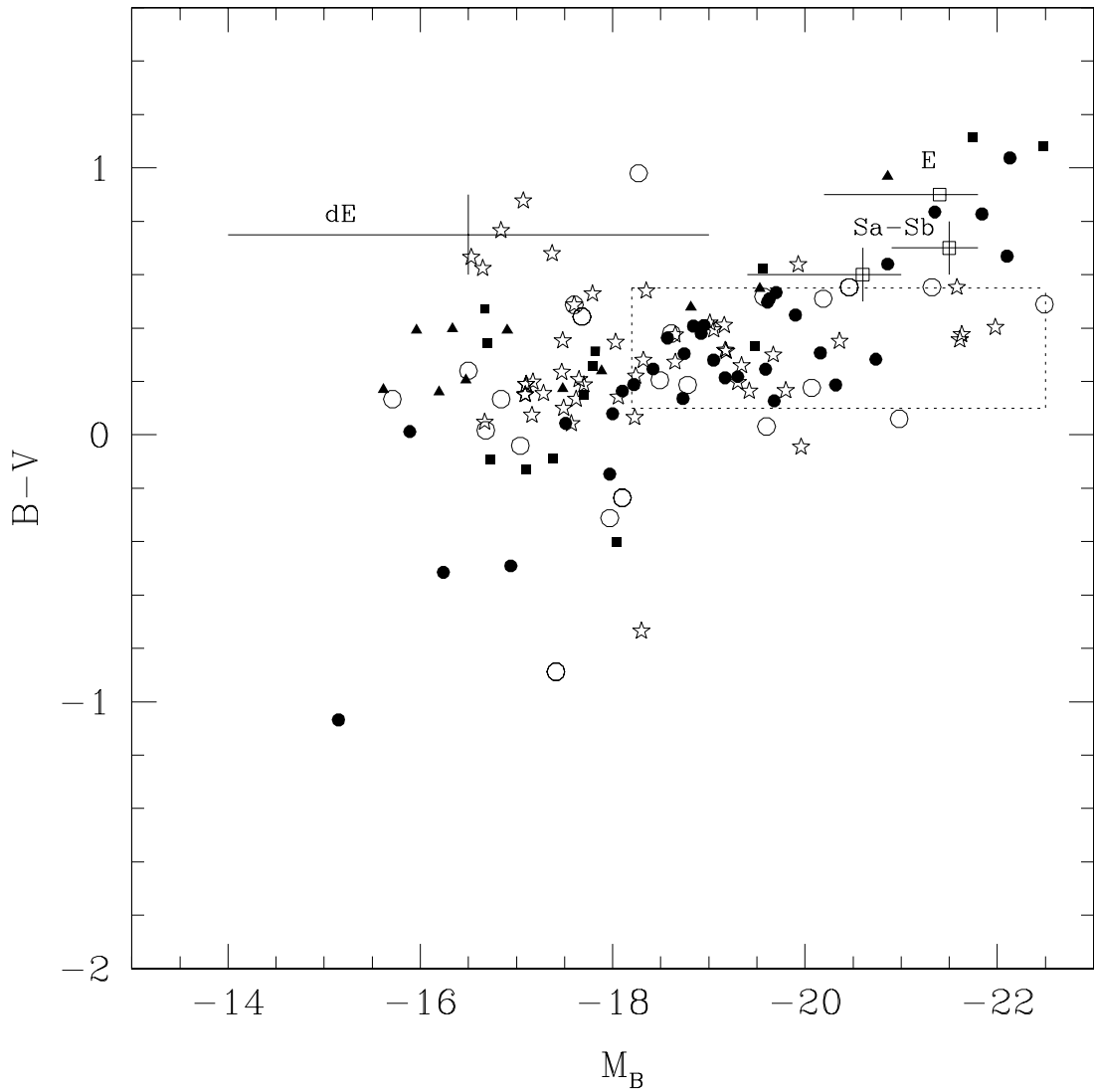}
\caption{Rest-frame B-V vs absolute B magnitude. 
The three empty squares are values typical of E, Sa-Sb, Sc-Irr (clockwise); 
the cross on the top left corresponds to the dE and dSph; the dotted region corresponds to the strong star-forming 
galaxies (Bershady et al. 2000) which contains blue nucleated galaxies, compact narrow emission-line galaxies and small, blue
galaxies at intermediate redshifts. Data symbols are: filled triangles=$0.2<z<0.4$, filled squares=$0.4<z<0.6$, 
filled circles=$0.6<z<0.8$, five pointed stars=$0.8<z<1.0$, empty circles=$1.0<z<1.2$.
\label{plotbbvall}}
\end{figure}

\begin{figure}
\plotone{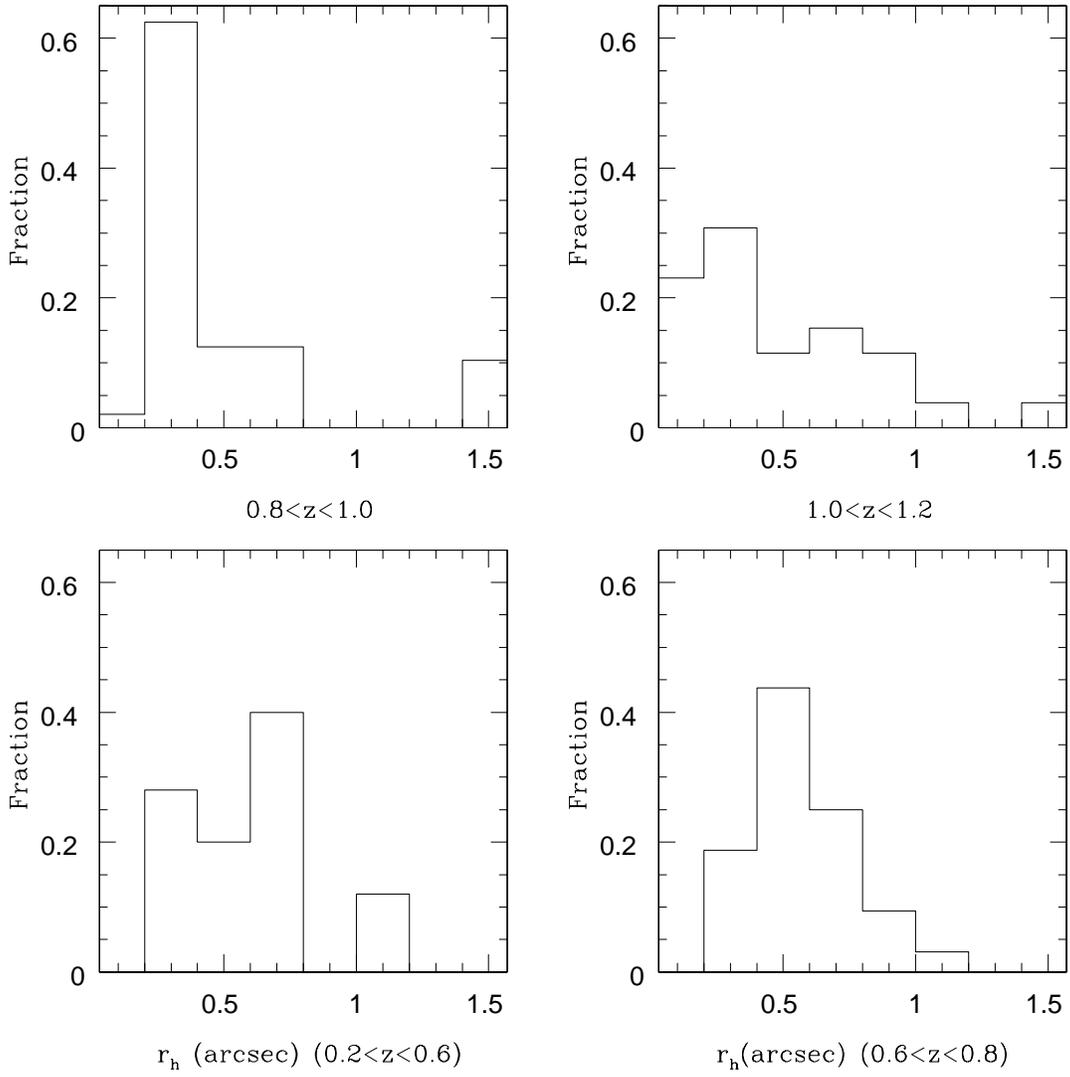}
\caption{Observed half-light radii (arcsec) measured in the rest-frame B band.
\label{histlightarcsec}}
\end{figure}

\begin{figure}
\plotone{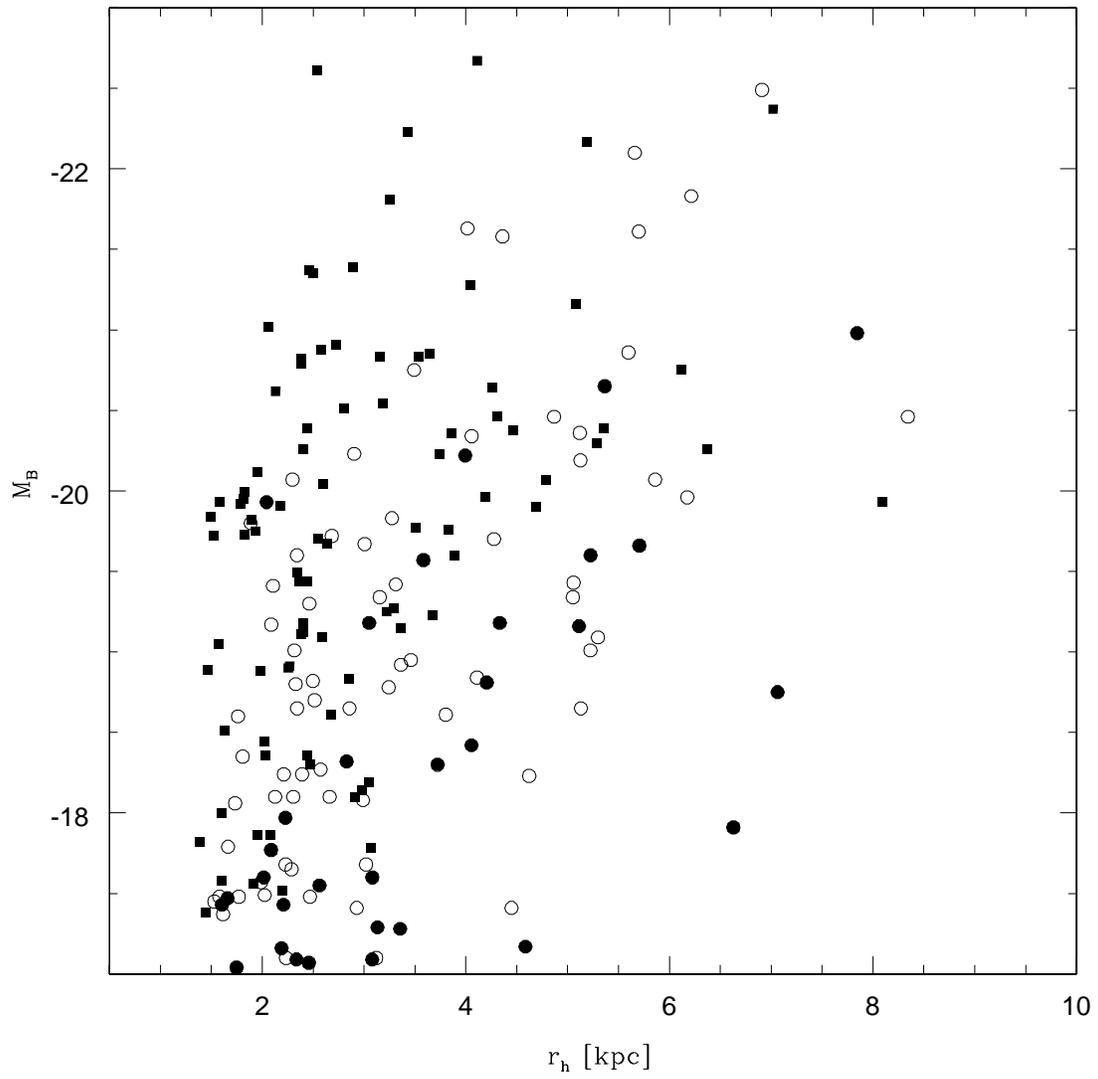}
\caption{Observed half-light radii (kpc) versus B absolute magnitude (circles)
at 0.7$<z<$1.4 (low surface brightness objects are shown in filled-cricles). 
The HDF-N data for the same redshift range from
Papovich et al. 2005 are shown in filled-squares. 
\label{plotblightkpchdf}}
\end{figure}

\begin{figure*}
\plotone{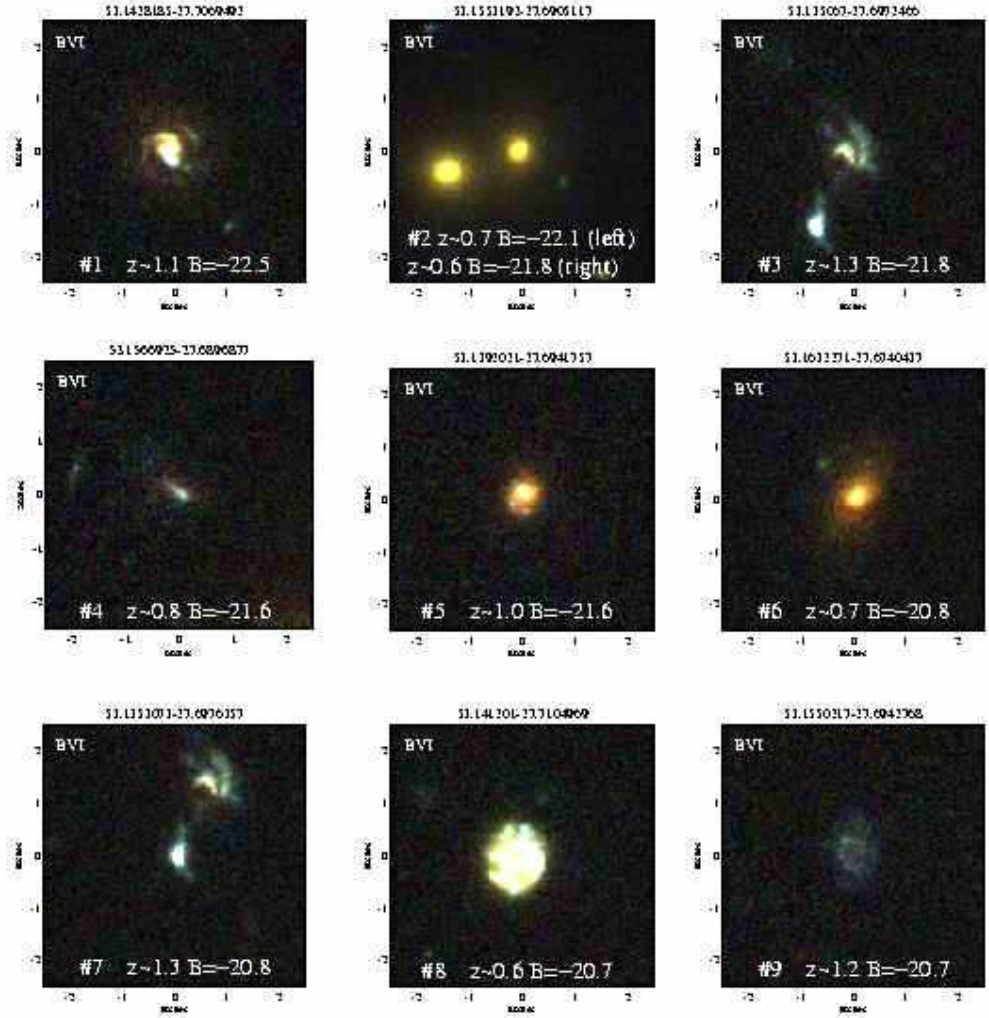}
\caption{A gallery of the most luminous objects (M$_{\rm B}$ $<$ --20.5) in the UV-selected sample. 
Each image is a 3-color (BVi) postage stamp montage  5$\times$5 arcsec$^{2}$. 
\label{luminous}}
\end{figure*}

We are grateful to G. Vazquez for providing us with the models and data 
used in Fig.\ref{plotvibvdatasb99} and to the GOODS team. 
Support for this work was provided by NASA through grants GO09583.01-96A and GO09481.01-A from 
the Space Telescope Science Institute, which is operated by the Association of Universities for Research 
in Astronomy, Inc., under NASA contract NAS5-26555.

\clearpage

\begin{thebibliography}{}
\bibitem[Abraham et al.(1996)]{ab96} Abraham, R.G., et al., 1996, MNRAS, 279, 47
\bibitem[Bell et al. (2004)]{be04} Bell, E. F., et al., 2004, ApJ, 608, 752
\bibitem[Bershady et al. (2000)]{ber00} Bershady, M. A., Jangren, A., \& Conselice, C. J. 2000, AJ, 119, 2645
\bibitem[Bertin et al.(1996)]{ber96} Bertin, E., \& Arnouts, S. 1996, \aaps, 117, 393
\bibitem[Bolzonella et al. (2000)]{bol00} Bolzonella, M., Miralles, J. 
-M., \& Pell\'{o}, R. 2000, A\&A, 363, 476
\bibitem[Casertano et al.(2000)]{2000AJ....120.2747C} Casertano, S., et 
al.\ 2000, \aj, 120, 2747 
\bibitem[Coleman et al.(1980)]{col80} Coleman, G. D., Wu, C.-C., \& 
Weedman, D. W. 1980, ApJS, 43, 393
\bibitem[Conselice (2003)]{Con03} Conselice, C. J. 2003, ApJS, 147, 1 
\bibitem[Corbin et al. (2003)]{Cor03} Corbin, M. R., Urban, A., Stobie, E., Thompson, R. I., \& Schneider,
G. 2001, ApJ, 551, 23 
\bibitem[Dahlen et al. (2005)]{dah05} Dahlen, T., et al. 2005, ApJ, 631, 126
\bibitem[Ferguson et al. (2004)]{fer04} Ferguson, H. C., et al., 2004, ApJ, L107
\bibitem[Giavalisco et al. (2004)]{gia04} Giavalisco, M., et al., 2004, 
ApJ, 600, L93
\bibitem[Heavens et al. (2004)]{Hea04} Heavens, A, Panter, B., Jimenez, R., \& Dunlop, J. 2004, Nature,
428, 625
\bibitem[Heckman et al. (2005)]{Hec05} Heckman, T. M., et al., 2005, ApJ, 619, 35 
\bibitem[Kinney et al. (1996)]{kin96} Kinney, A. L., Calzetti, D., 
Bohlin, R. C., McQuade, K., Storchi-Bergmann, T., \& Schmitt, H. R. 
1996, ApJ, 467, 38
\bibitem[Leitherer et al.(1999)]{lei99} Leitherer, C., et al. 1999, \apjs, 123, 3 (Starburst99)
\bibitem[Lotz et al. (2005)]{lot05} Lotz, J. M., Primack, J., \& Madau, P. 2005, AJ, 128, 163
\bibitem[Madau (1996)]{Mad96} Madau, P., et al. 1996, MNRAS, 283, 1388
\bibitem[Menanteau (2001)]{Men01} Menanteau, F., Abraham, R.G., \& Ellis, R.S. 2001, MNRAS, 322, 1
\bibitem[Mobasher (2004b)]{Mob04b} Mobasher, B., et al., 2004, ApJ, 600, L143
\bibitem[Papovich (2005)]{Pap05} Papovich, C., et al., 2005 ApJ accepted (astro-ph 0501088)
\bibitem[Pettini (2001)]{Pet01} Pettini, M., et al., 2001, ApJ, 554, 981
\bibitem[Steidel (1995)]{Ste95} Steidel, C. C., Pettini, M., \& Hamilton, D. 1995, AJ, 119, 2519
\bibitem[Stought (2002)]{Sto02} Stoughton, C., et al. 2002, AJ, 123, 485 (erratum 123, 3487)
\bibitem[Teplitz (2005)]{Tep05} Teplitz, H., et al. 2005 in preparation
\bibitem[van (1996)]{van96} van den Bergh, S., Abraham, R.G., Ellis, R.S., Tanvir, N.R., 
Santiago, B.X., Glazebrook, K.G., 1996, AJ, 112, 359
\bibitem[Vazquez (2005)]{Vaz05} Vazquez, G. A., Leitherer, C. 2005, ApJ, 621, 695
\bibitem[Wadadekar (2005)]{Wad05} Wadadekar et al. 2005, PASP, submitted
\bibitem[Whitmore (1999)]{Whi99} Whitmore, B. C., et al., 1999, AJ, 118, 1551
\bibitem[Wiegert (2004)]{Wie04} Wiegert, T., de Mello, D. F., \& Horellou, C. 2004, \aa, 426, 455
\end{thebibliography}
\end{document}